# Nano-patterning process based on epitaxial masking for the fabrication of electronic and spintronic devices made of $La_{0.67}Sr_{0.33}MnO_3/LaAlO_3/SrTiO_3$ heterostructures with *in situ* interfaces

F. Telesio, L. Pellegrino, I. Pallecchi [a], D. Marré

CNR-SPIN Genova and Università di Genova, Via Dodecaneso 33, 16146 Genova, Italy

E. Esposito

CNR-IMM, Via Pietro Castellino 111, 80131 Napoli, Italy

E. di Gennaro, Amit Khare [b], F. Miletto Granozio

CNR-SPIN Napoli and Dipartimento di Fisica, Università di Napoli "Federico II", 80126 Napoli, Italy

[a] Electronic mail: Ilaria.pallecchi@spin.cnr.it

[b] Present address: Sungkyunkwan University, Suwon 440 746, South Korea

The fabrication of oxide electronics devices is presently hindered by the lack of standardized and well established patterning procedures, applicable down to the nanoscale. In this work, we propose a procedure to obtain patterns with resolution around 100nm on $(La,Sr)MnO_3/LaAlO_3/SrTiO_3$ heterostructures. Our method is based on a multi-step technique, which includes wet and dry etching, epitaxial masking and e-beam lithography. Our procedure is devised to define independent patterns on the interfacial two dimensional electron gas and on the metallic top electrode, while preserving an all-*in-situ* approach for the heterostructure growth. We show results on nanoscale devices based on $(La,Sr)MnO_3/LaAlO_3/SrTiO_3$, suitable for oxide spintronics applications.

# I. INTRODUCTION

The discovery of a high mobility two-dimensional electron gas (2DEG) living at the interface between band insulators strontium titanate (SrTiO$_3$) and lanthanum aluminate (LaAlO$_3$) [1] has given a new impulse to research in oxide electronics. Such 2DEG forms as a consequence of the so called polar catastrophe, causing electronic reconstruction when a layer of LaAlO$_3$ of thickness equal or larger than a critical value of 4 unit cells (u.c.) [2] is deposited onto a TiO$_2$ terminated SrTiO$_3$ single crystal. This finding has suggested that, in perspective, these interfaces may be the building blocks for an oxide-based electronics analogous to the GaAs/AlGaAs semiconducting quantum wells, with the further benefit of possible new integrated functionalities. Indeed, in oxides, the tight interplay of charge, spin and orbital degrees of freedom gives rise to a variety of phenomena, such as Mott localization, charge, spin and orbital orderings, metal–insulator transitions, multiferroism and superconductivity. Further possibilities can be envisaged when such bulk properties are combined with interface mechanisms at the nanoscale [3], in the fabrication of electronic devices [4]. Remarkably, correlated oxides with high density of states have nanometric characteristic screening lengths and are thus suitable for the fabrication of nanoscale devices with high density of integration, overcoming major limits of conventional electronics.

Although patterning on the micron and nanoscale is the starting point and groundwork for this research field and its potential applications, the establishment of a standard protocol to pattern these system lags behind yet. This is mainly due to the difficulty of wet-etching LaAlO$_3$, as well as to the problems associated to dry etching of oxides. Indeed, it is well known that prolonged dry etching causes deoxygenation of the SrTiO$_3$ substrate, making it conducting, and suppression of LaAlO$_3$/SrTiO$_3$ conductivity, making it insulating, thus affecting dramatically the transport properties of the LaAlO$_3$/SrTiO$_3$ system and precluding device fabrication [5].

Several methods for patterning have been proposed to far. Some of them are based on the pre-patterning of an amorphous layer, that allows the formation of the 2DEG only in those selected areas where the epitaxial LaAlO$_3$/SrTiO$_3$ interface is deposited. Among these methods, the one proposed in ref. [6] starts from a pattern of amorphous SrTiO$_3$, followed by the deposition of crystalline LaAlO$_3$. The 2DEG forms only in the areas with no amorphous SrTiO$_3$ underneath, while the areas with buried amorphous SrTiO$_3$ remain insulating. The method proposed in ref. [7] starts by the deposition of 2 unit cells of crystalline LaAlO$_3$ on a SrTiO$_3$ crystalline substrate, followed by lithographic patterning, deposition of amorphous AlO$_x$ and lift off. Finally another layer of 2 or more unit cells of crystalline LaAlO$_3$ is deposited, so that the areas masked by AlO$_x$ turn out insulating because they have a subcritical LaAlO$_3$ thickness of 2 unit cells, while the areas not covered by the mask have a total LaAlO$_3$ thickness of 4 or more unit cells, larger than the threshold for 2DEG formation. Even if AlO$_x$ deposition and lift off are carried out *ex situ*, the active interface LaAlO$_3$/SrTiO$_3$ is grown *in situ* in the initial step. The method reported in ref. [8] starts from a patterned layer of amorphous AlO$_x$ on a SrTiO$_3$ crystalline substrate, followed by the deposition of crystalline LaAlO$_3$. A lift-off process driven by the AlO$_x$ layer in a 4M aqueous NaOH solution leaves a patterned 2DEG in the areas where no AlO$_x$ was present. Other kinds of methods to pattern the LaAlO$_3$/SrTiO$_3$ system are based on a dry etching step. Among these methods, the one reported in ref. [9], based on the use of electron beam lithography in combination with reactive ion etching, allows fabrication of LaAlO$_3$/SrTiO$_3$ channels of width down to 100 nm and can be directly extended to optical lithography, thus being easily transferred to industrial patterning processes. The study of ref. [5] demonstrates that by careful tuning of the beam energy and dose, argon ion beam irradiation drives the LaAlO$_3$/SrTiO$_3$ interface to an insulating state before eventually oxygen vacancies are created in the SrTiO$_3$ substrate. Hence, low dose irradiation in combination with e-beam lithography allows patterning LaAlO$_3$/SrTiO$_3$ channels of width down to 50 nm. Other methods have been proposed to obtain low dimensional structures, such as wires, but are not suitable for more complex device geometries. For example, LaAlO$_3$/SrTiO$_3$ wires of 50nm width have been obtained using a hard mask of an amorphous oxide deposited in two steps, with the direction of impinging species first perpendicular and then tilted with respect to the substrate plane, so that the final deposition of crystalline LaAlO$_3$ occurs directly on the surface of the SrTiO$_3$ crystalline substrate only in narrow grooves [10]. LaAlO$_3$/SrTiO$_3$ nanowires have been also fabricated at the edges of adjacent LaAlO$_3$ micrometric tiles having thickness of 1 and 3 unit cells, respectively, so that quantum wells hosting one dimensional electron gases are formed in correspondence of the thickness steps [11]. We also mention the self patterning process based on the chemical termination of the SrTiO$_3$ substrate controlled by high temperature annealing, which allows to obtain long-range ordered nanometric stripes of hundreds nm thickness [12]. Finally, by means of voltages applied to the conducting tip of an atomic force microscope (AFM) [13], local and reversible creation and erasure of conducting regions at the LaAlO$_3$/SrTiO$_3$ interface with resolution down to 3 nm is possible. Drawbacks of this method are unsuitability to large scale patterning and room temperature stability of the written patterns.

From an analysis of all the above methods, a lack of versatility emerges. In particular, a method should be developed which on one hand allows arbitrary device shape with features of different scales from mm to hundreds nm, and on the other hand includes the presence of additional conducting oxide overlayers whose pattern is independent from the pattern of the LaAlO$_3$/SrTiO$_3$, while preserving the *in situ* interface character. Such need may be indeed the case in the fabrication of charge or spin injecting electrodes in electronic and spintronic devices, as well as in the fabrication of complex epitaxial oxide heterostructures for multifunctional operation. Indeed, any conducting overlayer must be removed from the areas surrounding the devices and this should be done without breaking the condition of *in situ* deposition of all the interfaces of the heterostructure.

In this work, we propose a versatile multi-step patterning method based on epitaxial masking for wet etching, an approach that has been successfully tested on perovkite oxides [14,15]. Our process is a development of the method proposed in ref. [8] with lift off of amorphous AlO$_x$ and includes also dry and wet etching, optical and e-beam lithography. We demonstrate the possibility of fabricating nanoscale devices for oxide electronic and spintronic devices having *in situ* interfaces. Specifically, we realize devices based on the (La,Sr)MnO$_3$/LaAlO$_3$/SrTiO$_3$ heterostructure, where the (La,Sr)MnO$_3$ layer is an half metal ferromagnet with Curie temperature around room temperature. The choice of (La,Sr)MnO$_3$ as metallic electrode is due to its perfect lattice matching with SrTiO$_3$, which guarantees the all-in-situ growth of high quality epitaxial interfaces for device operation, as well as to its ferromagnetic character, which guarantees spin polarization of charge carriers close to 100% for spintronic applications.

## II. EXPERIMENTAL TECHNIQUES AND RESULTS

LaAlO$_3$, SrTiO$_3$ and (La,Sr)MnO$_3$ layers are deposited by pulsed laser deposition (PLD), as described in ref. [16,17]. The same deposition conditions, are adopted for the two layers, i.e.: 5x10$^{-2}$ mbar oxygen pressure, T = 740 °C substrate temperature, 1.5 J/cm$^2$ laser spot fluence, 1 Hz repetition rate. The selected oxygen pressure value is intermediate between those that are generally considered as optimal for LaAlO$_3$ and La$_{0.67}$Sr$_{0.33}$MnO$_3$ growth, and is intended to guarantee the good properties of both layers. Morphology and roughness of each layer are monitored *in situ* by RHEED (Reflection High Energy Electron Diffraction) diagnostics, either on unpatterned twin samples or, when possible, directly on the patterned samples. A further *ex situ* check was performed by AFM. Sputtering of an aluminum target is used for the deposition of amorphous aluminum oxide. Patterning of the samples down to the micron scale is performed by standard optical lithography, while sub-micron scale patterning is performed by e-beam lithography, using a Raith 150 system. Dry etching is carried out by ion milling and wet etching by dipping in HCl. The etching processes are calibrated and monitored by AFM. Transport measurements down to 4.2 K are carried out in a Physical Properties Measurement System (PPMS) by Quantum Design.

    a) We start from 7x7 mm$^2$ TiO$_2$ terminated SrTiO$_3$ crystalline substrates purchased from CrysTec GmbH. On these substrates we realize a micrometric photoresist pattern (see figure 1a). This pattern is realized

both with optical lithography using positive (Shipley 1813) or negative (ma-N 1407) photoresist or alternatively by e-beam lithography (PMMA). This pre-pattern allows to access each of the 9 nanometric devices with sub-mm size bonding pads. Moreover, it contains two sets of markers for each device for e-beam alignment and conducting paths to ground all the devices during e-beam exposure, thus avoiding charging effects. After exposure and development, the areas that are to become the devices remain covered with resist, while the surrounding areas are exposed to air.

b) We then deposit a 100-140 nm thick amorphous layer of $AlO_x$ (see figure 1b), by sputtering an aluminium target.

c) In the following step, a lift off process in acetone allows to obtain a pattern where the areas of the substrate surface that are to be the devices are exposed to air, while the surrounding areas are covered with amorphous $AlO_x$ (see figure 1c). Thanks to either the micrometric thickness of the positive photoresist or the undercut of the negative one, the resolution of the photoresist pattern is easily replicated after the lift-off process. Moreover, AFM imaging of the areas exposed to air demonstrates that atomically flat surfaces with one unit cell high terraces are preserved, indicating that the patterning prior to deposition does not affect the quality of the substrate.

d) At this point, the deposition of the heterostructure is carried out by pulsed laser ablation (see figure 1d). The stacking sequence of our heterostructure is $SrTiO_3$(12 u.c.)/(La,Sr)$MnO_3$(52 u.c.)/$LaAlO_3$(5 u.c.)/$SrTiO_3$(substrate).

From *in situ* RHEED monitoring, we see that the deposition of the LaAlO$_3$ layer, as well as most of the La$_{0.67}$Sr$_{0.33}$MnO$_3$ layer, occurs in a layer-by-layer two-dimensional mode (see figure 2). Deposition rates measured from RHEED oscillations are of the order of 18 shots/u.c. for LaAlO$_3$ and about 28 shots/u.c. for (La,Sr)MnO$_3$. The two-dimensional morphology of the first (La,Sr)MnO$_3$ layers is also confirmed by AFM imaging carried out on a similar heterostructure with thin (8 u.c.) La$_{0.67}$Sr$_{0.33}$MnO$_3$ layer, shown in figure 3a. As the La$_{0.67}$Sr$_{0.33}$MnO$_3$ layer deposition proceeds, three dimensional structures typical of island growth mode start to appear in RHEED patterns. Correspondingly, surface roughness of 0.6 - 4 nm is observed by AFM, as displayed in figure 3b. Such overgrowths and particulates may perforate the barrier of the top SrTiO$_3$ protecting layer, acting as a path for the acid solution producing pinholes on the La$_{0.67}$Sr$_{0.33}$MnO$_3$ layer. Such effect is visible in the AFM images displayed in figures 3c and 3d. In particular, figure 3c shows the effect of a prolonged HCl etching, to emphasize this effect. In actual devices, pinholes have lower density, as shown in figure 3d and in figure 5b, because the HCl etching is limited in time and residual PMMA acts as an additional protective barrier. The uppermost SrTiO$_3$ layer of our heterostructure is the epitaxial mask and its thickness is thus chosen on the basis of (i) transparency of (La,Sr)MnO$_3$ pattern for sufficient contrast in e-beam imaging, (ii) proper working as epitaxial mask for wet etching, thus it must be well connected and fully cover the underlying heterostructure, (iii) small enough thickness so that it is fully removed in the dry etching step for a dry etching time that does not simultaneously remove the whole PMMA thickness, as better explained later on in points f) and g). Hence, requirements (i) and (iii) sets the upper limits to the cap-layer thickness, while (ii) sets the lower limit. The latter lower limit also depends on the roughness of the heterostructure, which in our case is 2-4 nm, as measured by AFM over areas of few squared micron.

e) In the next step, lift off of the $AlO_x$ layer is carried out in 4 M aqueous NaOH solution heated at 65 °C in ultrasonic bath (see figure 1e). This step takes several hours to be completed and it is favoured by the granularity of the $AlO_x$ layer, which is visible in the AFM image of figure 4.

f) Then, PMMA 2% is spinned at 4500 r.p.m. (revolutions per minute) for 45 sec. on the sample (see figure 1f). The resulting PMMA thickness of 80 nm is chosen as a compromise between e-beam resolution and the condition that the duration of the following dry etching, necessary to remove the $SrTiO_3$ cap layer, is short enough to prevent a complete removal of the PMMA layer. Indeed, the areas covered by PMMA must have a well-connected and undamaged epitaxial $SrTiO_3$ cap layer that should properly protect the underlying heterostructure from wet etching.

g) The PMMA on the sample is exposed to the beam for electron beam lithography, according to the desired nano-pattern (see figure 1g). In our case, the prototype device is a planar spin valve with ferromagnetic $La_{0.67}Sr_{0.33}MnO_3$ electrodes of different shapes that inject spin polarized electrons in the LaAlO$_3$/SrTiO$_3$ 2DEG located in the gap between the electrodes themselves. The thin electrode is 0.5 μm wide and 5 μm long and the gap between the electrodes is ~150 nm wide. The e-beam exposure parameters are chosen such that they provide a suitable resist edge profile for the dry-etching post process. A second flood exposure is then performed to improve the PMMA dry etching resistance. Finally the PMMA is developed.

h) The sample undergoes a dry etching (see figure 1h) for 100 s, with Ar ions of energy 500 eV and current of 10 mA, corresponding to a current density on the sample of 0.25 mA/cm$^2$. This time is long enough to remove completely the SrTiO$_3$ cap layer from the exposed areas but short enough to let the cap layer protected by PMMA remain in its whole integrity, and be used as a mask for the following wet etching step. Most importantly, the dry etching time and energy are soft enough to prevent deoxygenation of the 2DEG interface, which is buried deep into the heterostructure and is therefore not directly targeted by the impinging ions. According to an estimation made using a phenomenological formula as in ref. [18] and to calculations made by Monte Carlo method using the "Stopping and range of ions in matter" (SRIM) software [19], we expect the penetration depth of these impinging ions through either (La,Sr)MnO$_3$ or underlying SrTiO$_3$ to be 3-5 nm. Although neither the phenomenological formula nor the simulation take into account possible migration of oxygen vacancies created in the damaged surface layer, we do not believe this effect to be significant, given that the damaged surface layer is 4-6 times thinner than the (La,Sr)MnO$_3$ layer successively removed by wet etching.

i) The La$_{0.67}$Sr$_{0.33}$MnO$_3$ layer is now partially protected by the epitaxial cap layer and partially exposed to air. As a next step, a wet etching in 3.7% HCl is carried out for few tens of seconds (see figure 1i). Our epitaxial masking approach is made possible by two facts: i) that SrTiO$_3$ is not etched by HCl, while La$_{0.67}$Sr$_{0.33}$MnO$_3$ is [14,15]; ii) that the epitaxial cap layer adheres so perfectly to the underlying heterostructure to guarantee a reliable protection [14], unlike PMMA nanostructures on oxide films that hardly sustain prolonged wet etching in HCl. Such preparation step can be extended to different systems, since any other isostructural oxide that can be selectively etched with respect to any other oxide epitaxial cap layer, can be deposited and patterned in the same way on the top of the LaAlO$_3$/SrTiO$_3$ interface.

The patterning process is concluded. In figure 5 we present an optical image of a device and an AFM image of the active area of the same device. In general, features with resolution down to 100 nm are obtained after the overall process.

As this paper is focused on the preparation process, the characterization of the final spintronic device will be addressed elsewhere. However, we mention that in the case that planar resistance in the La$_{0.67}$Sr$_{0.33}$MnO$_3$ electrodes overwhelms the resistance across the LaAlO$_3$/SrTiO$_3$ interface between the electrodes, a 15 nm thick overlayer of gold can be deposited on the top of the whole heterostructure and patterned simultaneously with the SrTiO$_3$ cap layer by dry etching (note that the dry etching rate for Au is much larger than the one of SrTiO$_3$, hence an additional 15 nm thick overlayer of gold does not require to reconsider dry etching times discussed in point h)). The gold layer acts as an electrical shunt for the in-plane current path along the La$_{0.67}$Sr$_{0.33}$MnO$_3$ electrodes, allowing direct measurement of the LaAlO$_3$/SrTiO$_3$ interface between the electrodes. Indeed, with a suitable voltage bias, the electric carriers travel in the gold layer along the planar segment of their path, then they tunnel from gold to La$_{0.67}$Sr$_{0.33}$MnO$_3$ across the SrTiO$_3$ cap layer and finally they are then injected directly in the LaAlO$_3$/SrTiO$_3$ interface between the electrodes.

On the other hand, in this context our focus is demonstrating that the whole patterning process does not damage the transport properties of the La$_{0.67}$Sr$_{0.33}$MnO$_3$ and LaAlO$_3$/SrTiO$_3$ layers. In figure 6a), we show resistivity versus temperature behavior of a device where the two (La,Sr)MnO$_3$ electrodes are connected (the gap between them is zero), so that the transport mainly probes the (La,Sr)MnO$_3$ layer. It can be seen that the sample exhibits a metal-insulator transition around room temperature, characteristic of high quality (La,Sr)MnO$_3$ and also the resistivity value in the range of $10^{-4}$ $\Omega$m is consistent with typical values for this compound. In figure 6b), the sheet resistance curve measured on the LaAlO$_3$/SrTiO$_3$ layer is shown, after complete removal of the overlying layers (La$_{0.67}$Sr$_{0.33}$MnO$_3$ and SrTiO$_3$ cap layer), following the procedures h)-i). Also in this case, we find the typical behavior of LaAlO$_3$/SrTiO$_3$ 2DEGs, with room temperature sheet resistance values in the tens of k$\Omega$ range and metallic behavior down to low temperature. Hence, we can conclude that the heterostructure properties are not affected throughout the patterning process.

## III. CONCLUSIONS

In this work we propose a patterning multi-step procedure based on wet and dry etching, epitaxial masking and e-beam lithography, that allows to obtain patterns on LaAlO$_3$/SrTiO$_3$ interfaces with additional *in situ* conducting overlayers. The procedure is particularly versatile in terms of device architecture, as compared to the methods proposed so far, and does not damage the transport properties of the heterostructure layers. Patterns with resolution around 100 nm are obtained.

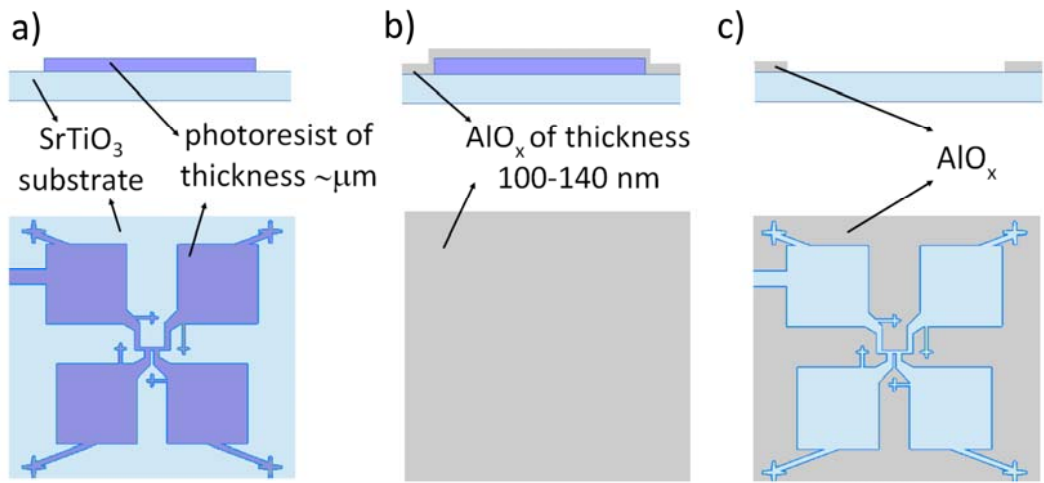
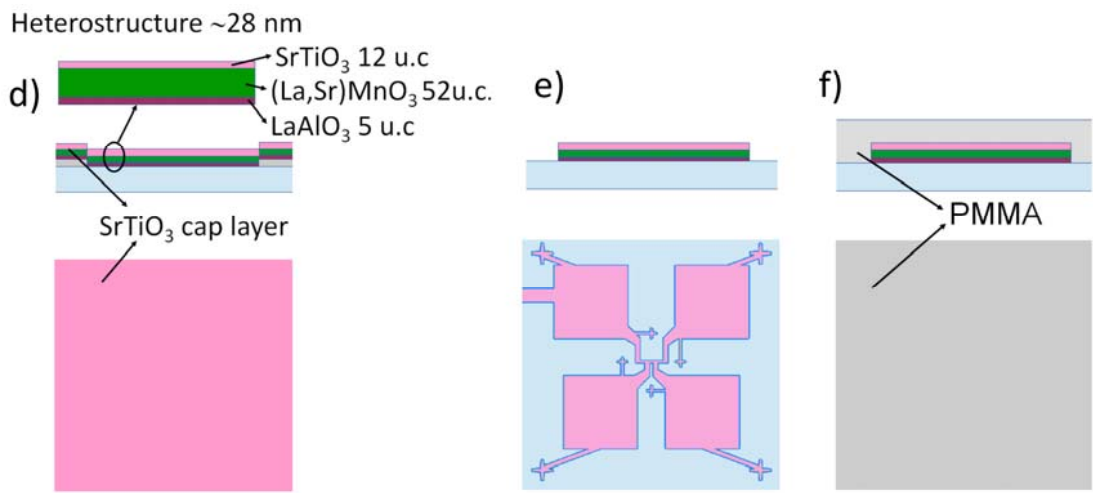
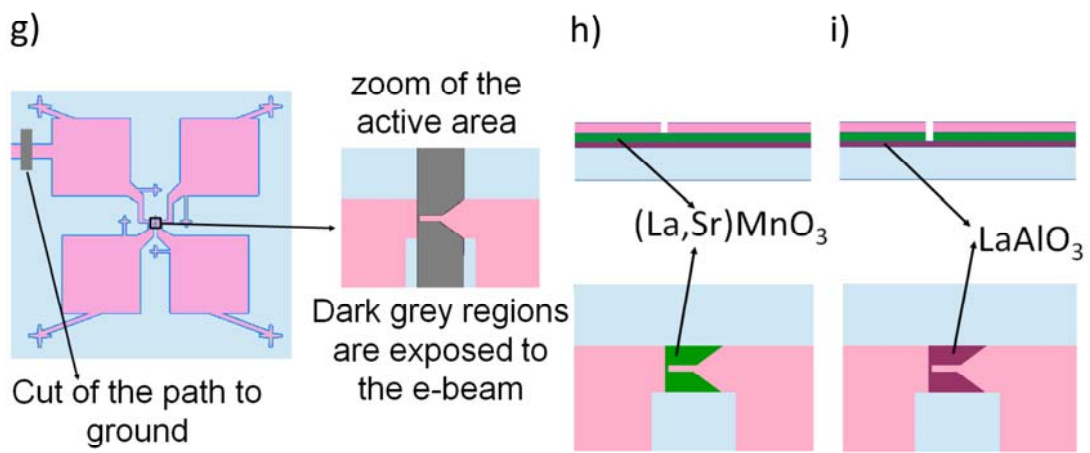

FIG. 1. (Color online) Sketch of the multistep patterning process; panels a) to i) depict the steps described in the text at the corresponding a) to i) points, namely a) photoresist pre-patterning, b) AlO$_x$ deposition, c) AlOx lift-off, d) heterostructure deposition, e) heterostructure lift-off, f) PMMA spinning, g) e-beam patterning, h) dry etching of the SrTiO$_3$ cap layer, i) HCL wet etching of the La$_{0.67}$Sr$_{0.33}$MnO$_3$ layer. Both top views and side views are shown. Top views of a single device are in panels a) to g), side views of a single device are in panels a) to f), top views of the active area of a device are in panels g) to i) and side views of the active area of a device are in panels h) and i).

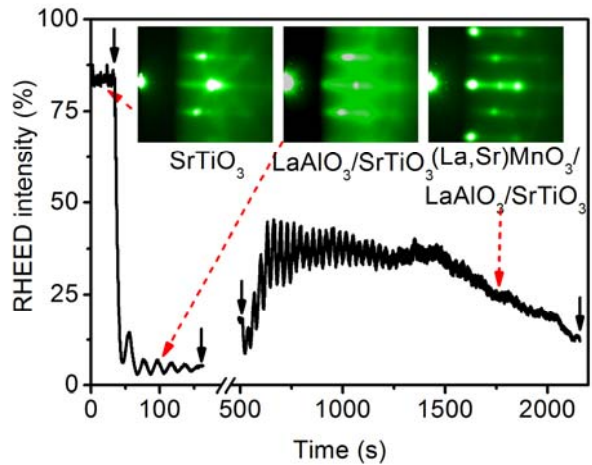

Figure 2: (color online) RHEED oscillations observed on a patterned heterostructure during the growth. The black arrows indicate the start and the stop time for the ablation processes. In the insets, RHEED patterns acquired at different times, indicated by red arrows, are shown.

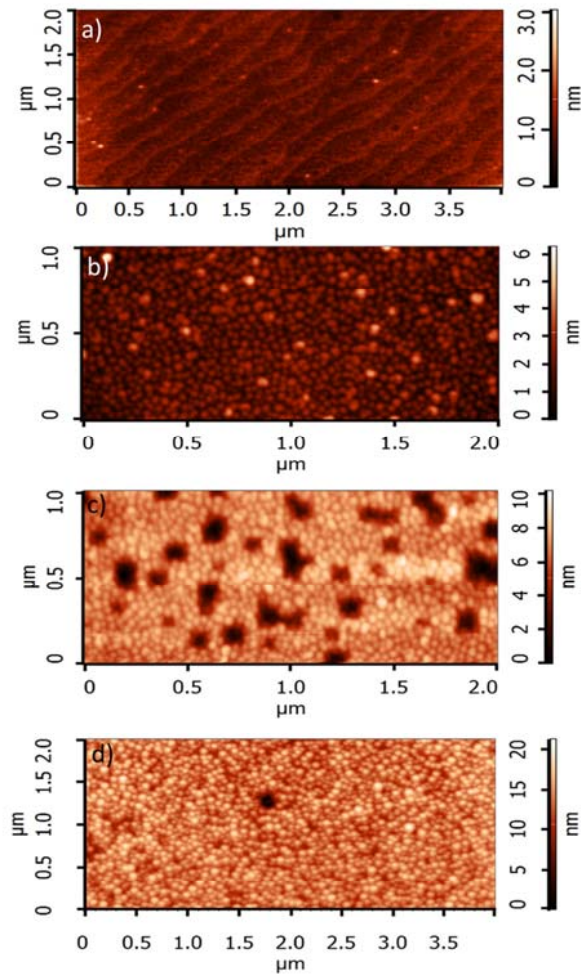

Figure 3: (color online) a) AFM image of a (La,Sr)MnO$_3$(8 u.c.)/LaAlO$_3$(5 u.c.)/SrTiO$_3$(substrate) heterostructure. Terraces are clearly visible on the (La,Sr)MnO$_3$ surface. b) AFM image of a complete SrTiO$_3$(12 u.c.)/ (La,Sr)MnO$_3$(52 u.c.)/LaAlO$_3$(5 u.c.)/SrTiO$_3$(substrate) heterostructure. Particulates are visible, determining a r.m.s. roughness of 0.6 nm. c) AFM image of the same heterostructure after prolonged HCl etching, showing pinholes originating from former particulates. d) AFM image of the surface of a final device, showing a smaller density of pinholes.

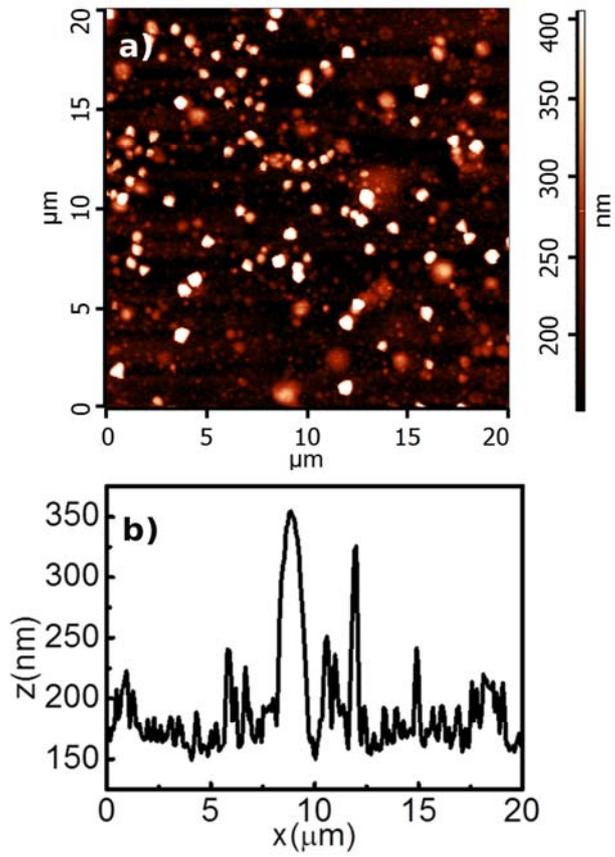

Figure 4: (color online) AFM image (a) and scanline (b) of 100nm amorphous $AlO_x$ deposited by sputtering. The roughness is 62nm rms.

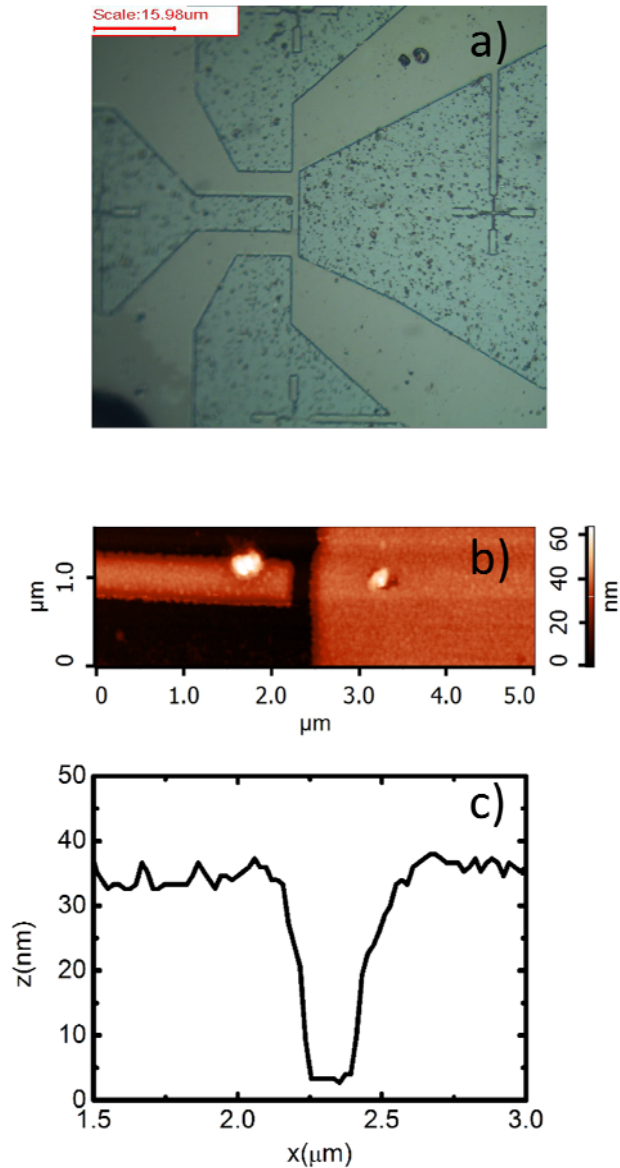

Figure 5: (color online) a) Optical image of a finished spin injection device is shown. The areas with high density of particulates are those with residual particulates from the lift-off of amorphous $AlO_x$ and overlayers from the PLD deposition. b) AFM image of the active area of a device and c) partial line scan across the gap of the device are shown. In this pattern, the gap is around 220 nm and the electrode width is 550 nm.

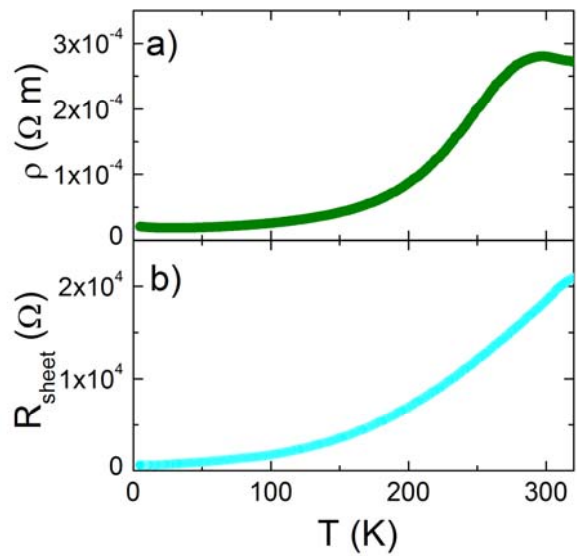

Figure 6: (color online) a) Resistivity versus temperature behavior of the La$_{0.67}$Sr$_{0.33}$MnO$_3$ electrodes of a device, where the two La$_{0.67}$Sr$_{0.33}$MnO$_3$ electrodes are connected (the gap between them is zero). b) Sheet resistance versus temperature behavior of the LaAlO$_3$/SrTiO$_3$ 2DEG, measured after removing the overlying layers.

[1] A. Ohtomo, and H. Y. Hwang, Nature **427**, 423-426 (2004).

[2] S. Thiel, G. Hammerl, A. Schmehl, C. W. Schneider, and J. Mannhart, Science **313**, 1942 (2006).

[3] J. A. Sulpizio, S. Ilani, P. Irvin, and J. Levy, Annual Review of Materials Research **44**, 117-149 (2014).

[4]D. F. Bogorin, P. Irvin, C. Cen, and J. Levy, chapter for the book "Multifunctional Oxide Heterostructures", edited by E. Y. Tsymbal, E. Dagotto, C. B. Eom, and R. Ramesh (Oxford University Press, 2010), also available at arXiv:1011.5290.

[5]P. P. Aurino, A. Kalabukhov, N. Tuzla, E. Olsson, T. Claeson, and D. Winkler, Appl. Phys. Lett. **102**, 201610 (2013).

[6]D. Stornaiuolo, S. Gariglio, N. J. G. Couto, A. Fête, A. D. Caviglia, G. Seyfarth, D. Jaccard, A. F. Morpurgo, and J.-M.Triscone, Appl. Phys. Lett. **101**, 222601 (2012).

[7]C. Schneider, S. Thiel, G. Hammerl, C. Richter, and J. Mannhart, Appl. Phys. Lett. **89**, 122101 (2006).

[8]N. Banerjee, M. Huijben, G. Koster, and G. Rijnders, Appl. Phys. Lett. **100**, 041601 (2012).

[9]M. Z. Minhas, H. H. Blaschek, F. Heyroth, and G. Schmidt, arXiv: 1502.06382.

[10]A. Ron, E. Maniv, D. Graf, J. -H. Park, and Y. Dagan, Phys. Rev. Lett. **113**, 216801 (2014).

[11]A. Ron, and Y. Dagan, Phys. Rev. Lett. **112**, 136801 (2014).

[12]M. Foerster, R. Bachelet, V. Laukhin, J. Fontcuberta, G. Herranz, and F. Sánchez, Appl. Phys. Lett. **100**, 231607 (2012).

[13]C. Cen, S. Thiel, G. Hammerl, C. W. Schneider, K. E. Andersen, C. S. Hellberg, J. Mannhart, and J. Levy, Nature Mat. **7**, 298 (2008).

[14]L. Pellegrino, I. Pallecchi, E. Bellingeri, G. Canu, A. S. Siri , D. Marré, Y. Yanagisawa, M. Ishikawa, T. Matsumoto, H. Tanaka, and T. Kawai, J. Nanosci. Nanotechnol. **10**, 4471 (2010).

[15]L. Pellegrino, M. Biasotti, E. Bellingeri, C. Bernini, A. S. Siri, and D. Marré, Adv. Mater. **21**, 2377 (2009).

[16]L. Parlato, R. Arpaia, C. De Lisio, F. Miletto Granozio, G. P. Pepe, P. Perna, V. Pagliarulo, C. Bonavolontà, M. Radovic, Y. Wang, Roman Sobolewski, and U. Scotti di Uccio Phys. Rev. B **87**, 134514 (2013) .

[17]C. Aruta, S. Amoruso, R. Bruzzese, X. Wang, D. Maccariello, F. Miletto Granozio, and U. Scotti di Uccio, Appl. Phys. Lett. **97**, 252105 (2010).

[18]D. W. Reagor, and V. Y. Butko, Nature Mat. **4**, 593 (2005).

[19]J. Ziegler, J. Biersack, and U. Littmark, The stopping of ions in matter (Pergamon, New York, 1985).